\documentclass[amsmath,amssymb,10pt,aps,notitlepage,prl,twocolumn,superscriptaddress,nofootinbib,longbibliography]{revtex4-1} 
 \usepackage[colorlinks=true]{hyperref}
 \usepackage{lipsum}
 \usepackage{graphicx}
\usepackage{latexsym}
\usepackage{amsmath}
\usepackage{amsthm}
\usepackage{amssymb}
\usepackage{epstopdf} 
\usepackage{enumitem}
\usepackage{setspace}
\usepackage{dcolumn}
\usepackage{bm}
\usepackage{setspace} 
\usepackage{slashed}
\usepackage{color}
\usepackage{youngtab}
\usepackage{tikz}
\usepackage{tikz-3dplot}
\usepackage{braket}
\usepackage{cancel}
\usepackage{subfigure}
\usepackage{pbox}
\usepackage[normalem]{ulem}

\usepackage{lipsum}
\graphicspath{{Figures/}}

\usetikzlibrary{shapes,snakes,arrows,chains,matrix,positioning,scopes,calc}
\usetikzlibrary{decorations.markings}

 \usepackage{moreverb}
\allowdisplaybreaks 
\begin{document}

\title{  Advancing Hybrid Quantum-Classical Algorithms via Mean-Operators} 
  
\author{Donggyu Kim}
 \affiliation{Department of Physics, Korea Advanced Institute of Science and Technology (KAIST), Daejeon 34141, Korea}

\author{Pureum Noh}
 \affiliation{Department of Physics, Korea Advanced Institute of Science and Technology (KAIST), Daejeon 34141, Korea}

\author{Hyun-Yong Lee}
\thanks{ hyunyong@korea.ac.kr}
\affiliation{Department of Applied Physics, Graduate School, Korea University, Sejong 30019, Korea}
\affiliation{Division of Display and Semiconductor Physics, Korea University, Sejong 30019, Korea}
\affiliation{Interdisciplinary Program in E$\cdot$ICT-Culture-Sports Convergence, Korea University, Sejong 30019, Korea}

 \author{Eun-Gook Moon}
\thanks{egmoon@kaist.ac.kr}
\affiliation{Department of Physics, Korea Advanced Institute of Science and Technology (KAIST), Daejeon 34141, Korea}

\date{\today}
\begin{abstract}
Entanglement in quantum many-body systems is the key concept for future technology and science, opening up a possibility to explore uncharted realms in an enormously large Hilbert space. The hybrid quantum-classical algorithms have been suggested to control quantum entanglement of many-body systems, and yet their applicability is intrinsically limited by the numbers of qubits and quantum operations. Here we propose a theory which overcomes the limitations by combining advantages of the hybrid algorithms and the standard mean-field-theory in condensed matter physics, named as mean-operator-theory. We demonstrate that the number of quantum operations to prepare an entangled target many-body state such as symmetry-protected-topological states is significantly reduced by introducing a mean-operator. We also show that a class of mean-operators is expressed as time-evolution operators and our theory is directly applicable to quantum simulations with $^{87}$Rb neutral atoms or trapped $^{40}$Ca$^+$ ions.
 \end{abstract}

\maketitle

{\it Introduction : } 
Remarkable quantum entanglement phenomena are uncovered by experimental advances in quantum science and technology.
Systems with about fifty qubits have been manipulated, and quantum entanglement achieves computational tasks well beyond classical computations as known as quantum supermacy \cite{Preskill18, Arute19}. 
Enormously entangled many-body states such as topologically ordered states have been realized in a 31-qubit superconducting quantum processor \cite{Google2021} and a 219-atom programmable quantum simulator of Rydberg atom arrays \cite{Semeghini2021} recently. Signatures of fractionalized particles also have been reported in systems with about the Avogadro number of electrons \cite{Zhou, Balents, Moessner2019, Takagi2019}, which manifests massive entanglement of many-body systems.

Theoretical efforts also have been made  to utilize quantum entanglement, and the variational hybrid quantum-classical algorithms such as the quantum-approximate-optimization-algorithm\,(QAOA) have been proposed \cite{Farhi14, McClean16}. The main idea is to employ classical computations to complement expensive quantum operations and design a variational circuit to prepare a target quantum state. Applicability of the algorithm is already demonstrated in trapped ion experiments \cite{Moll18, Britton12, Zoller19}.
However, the numbers of qubits and quantum operations are seriously limited in NISQ technology.
Recent theoretical works even find a fundamental limitation of QAOA that a deep circuit proportional to a system size is necessary to achieve a non-trivial quantum state \cite{Hsieh2019, Tang}. 
Thus, a theory, which reduces the number of quantum operations and applies to a system with an arbitrary number of qubits, is in great demand.  
 
\begin{figure}[tb]
\centering
\includegraphics[width=\columnwidth]{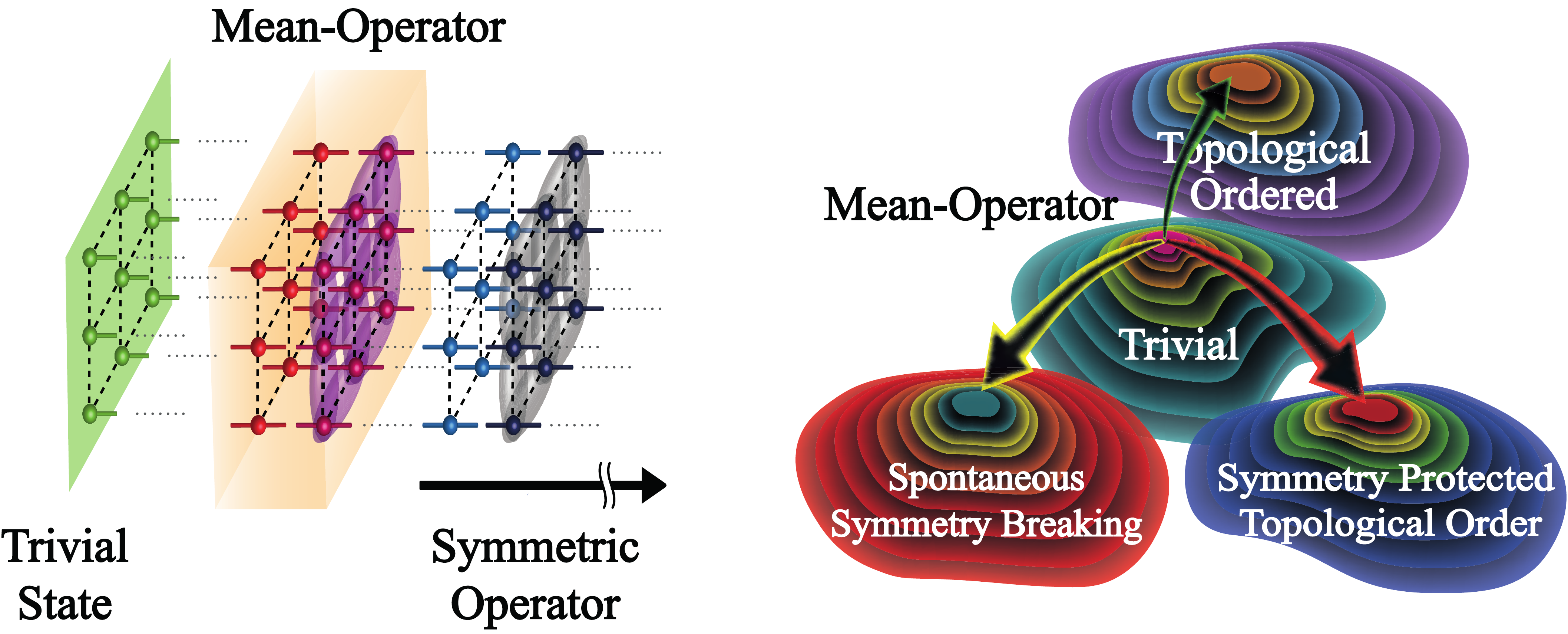} 
 \caption{ (a) Construction of a variational state by acting mean-operator and symmetric operator on a trivial state. (b) The mean-operators directly access the non-trivial phases such as the symmetry breaking, the symmetry protected topological and/or topologically ordered phases. The symmetric operator increases quantum entanglement, and its action is analogous to the standard hybrid quantum-classical algorithm such as QAOA.
  }
 \label{fig:0}
\end{figure}

In this work, we provide a variational theory which overcomes the limitations by combining advantages of the hybrid algorithms and the standard mean-field-theory\,(MFT), named as mean-operator-theory (MOT). 
Our main strategy is to let an operator with judicious guess, called a mean-operator, to capture essential characteristics of an entangled many-body state such as symmetry properties and entanglement patterns. 
And then, the conventional QAOA is performed to refine a variational state quantitatively as illustrated in Fig 1. The mean-operator plays a similar role of  a mean-field order parameter in the standard MFT. 
To find an explicit form of a mean-operator, it is particularly useful to employ intuitions from quantum many-body states in condensed matter such as symmetry and entanglement properties.  

We demonstrate that our theory significantly reduces the number of quantum operations to prepare target states and is even applicable to spatial dimensions higher than two. 
Furthermore, a class of mean-operators may be realized as time-evolution operators because they are chosen to be unitary operators, and thus our theory may be directly applicable to experiments of near-term quantum simulations including Rydberg atoms and trapped ions. 
Our theory may also serve as a theoretical tool to investigate general quantum many-body problems outperforming conventional mean-field type methods.

{\it Mean-Operator-Theory : } 
The MOT adopts the variational method of quantum mechanics and an ansatz state  of a system with a symmetry group  $\mathbb{G}$, 
\begin{eqnarray}
| \Psi ; \{\alpha, \beta\}_p, \{ \phi \} \rangle = \hat{S}(\{\alpha, \beta\}_p) \cdot \hat{M}(\{ \phi \}) \cdot | 0 \rangle, 
\end{eqnarray}
consists of the three main objects, 
\begin{itemize}
\item  $\hat{M}(\{ \phi \})$ : Mean-operator with a set of parameters $\{\phi\}$, which is non-trivial under $\mathbb{G}$ for a generic $ \phi $
\item $\hat{S}(\{\alpha, \beta\}_p)$ : Symmetric operator with a set of parameters $\{\alpha, \beta\}_p$, which is trivial under $\mathbb{G}$
\item $| 0 \rangle = \prod_j | +\rangle_j$ : Symmetric product state with a site index $j$, where $|+\rangle_j$ is trivial under $\mathbb{G}$.
\end{itemize}
Here, $p$ is an integer determining the complexity and accuracy of the ansatz and is referred to as the depth level of MOT. In this work, we assume the translational symmetry and that the representation of $\mathbb{G}$ is decomposed into the on-site unitary. 
By minimizing the energy expectation value of a Hamiltonian $H$, $\langle \Psi | H | \Psi \rangle / \langle \Psi | \Psi \rangle$, a better approximated ground state is obtained with more parameters.

 A symmetric operator, $\hat{S}(\{\alpha, \beta\}_p)$, employs the hybrid quantum-classical algorithms such as QAOA \cite{McClean16} and refines a variational state quantitatively. As usual, the total Hamiltonian is decomposed as $H=H_1+H_2$, where $H_1$ and $H_2$ do not commute each other, and the symmetric operator may be written as $U_S(\{\alpha, \beta\}_p)= \prod_{a=1}^p e^{-i \alpha_a H_1} e^{-i \beta_a H_2}$ with parameters $( \alpha_a, \beta_a)$ with a certain depth $p$. By construction, $H_1$ and $H_2$ are symmetric under $\mathbb{G}$.  

A mean-operator, $\hat{M}(\{ \phi \})$, significantly reduces the number of quantum operations to access non-trivial states, including topological or symmetry-broken states. 
The non-trivial property under $\mathbb{G}$  allows us to explore  a much wider region in the Hilbert space effectively than the standard QAOA.
In particular, there exists a special subset of $\{\phi_s\}$ which makes the variational state $\hat{M}(\{\phi_s\})|0\rangle$ both entangled and symmetric. 
It is obvious that the state is hard to be constructed by using the conventional QAOA.
In what follows, we apply our MOT to three different cases.

\begin{figure}[t]
\includegraphics[width=0.5\textwidth]{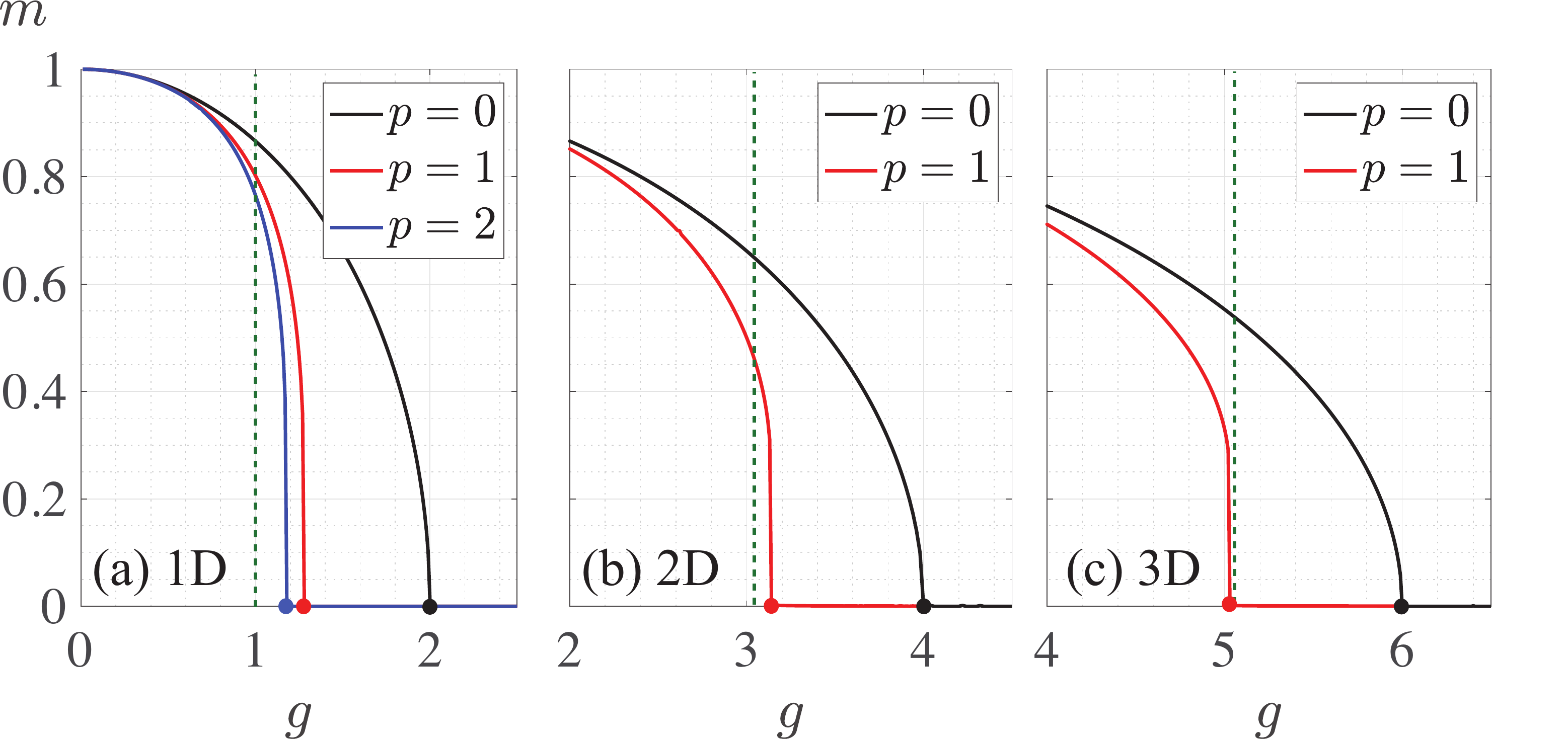} 
 	\caption{ Magnetization obtained by MOT in (a) one-dimensional, (b) two-dimensional and (c) three-dimensional transverse-field-Ising model, $H_1(g)$. Here, the green dotted lines stand for the exact critical points, i.e., $g_c^{\rm exact} = 1$ in 1D, $g^{\rm exact}_c = 3.044$ in 2D, and $g^{\rm CMC}_c = 5.158$ in 3D \cite{Blote02}. The depth-two MOT in 1D finds $g_c^{(2)} \simeq 1.18$, and the depth-one MOT finds $g_c^{(1)} \simeq 3.14$ and $g_c^{(1)} \simeq 5.03$ for 2D and 3D, respectively. }
 \label{fig:ssb}
\end{figure}

{\it Case 1) Spontaneously-symmetry-broken state : } 
Let us first consider spontaneous-symmetry-breaking. To be specific, we consider the transverse-field-Ising model \cite{Sachdev2011,Wen_QI} with a dimensionless coupling constant, $g$, 
\begin{eqnarray}
H_{1} (g)= - \sum_{\langle i,j \rangle} Z_i Z_j -g \sum_j X_j , 
\label{eq:tfi_model}
\end{eqnarray}
where Pauli matrices $X_j,Y_j,Z_j$ at site $j$ on a hypercubic lattice are used. The exchange energy scale is set to be unity, and the Hamiltonian enjoys the $\mathbb{Z}_2$ symmetry with $\{ I, \prod_j X_j \}$. The identity operator, $I$, is introduced, and the symmetric product state, $ |0\rangle =\prod_j |+\rangle_j$ with $X_j |+\rangle_j = |+\rangle_j$, is the ground state of $H_1(\infty)$.  

We consider the operators, 
\begin{eqnarray} 
&&\hat{M}_1(\phi ) = e^{-i \sum_j \phi Y_j} \nonumber \\
&&\hat{S}_1(\{\alpha, \beta\}_p) = \prod_{a=1}^p e^{-i \alpha_a \sum_j X_j} e^{-i \beta_a  \sum_{\langle i,j \rangle} Z_i Z_j}.  
\end{eqnarray}
For the depth-zero\,($p=0$), the expectation values of the order parameter and energy density are
$m(\phi) = \langle \Psi | Z_j| \Psi \rangle  =  -S_{2\phi}$ and $\epsilon(\phi) \equiv \langle  \Psi | H |   \Psi \rangle/ N =  -S_{2\phi}^2 - g C_{2\phi}$
with the total site number $N$. The short-handed notations, $C_{\theta} \equiv \cos(\theta)$ and  $S_{\theta} \equiv \sin(\theta)$, are used hereafter.
Minimizing the energy density determines the order parameter,
$m^{(0)} = \pm \sqrt{1-\left( \frac{g}{2d} \right)^2 }$, in spatial dimension $d$ 
, which gives the critical coupling constant, $g_{c}^{(0)}=2d$.    
Applying $\hat{S}_1$ with a non-zero $p$, systematic improvements are achieved. Fig.\,\ref{fig:ssb} presents the order parameter obtained by MOT for $d=1,2$ and $3$, of which the critical points are significantly improved with the non-zero depth.
Details on the computation of $\epsilon$ and $m$ are provided in Supplemental Materials\,(SM).

Few remarks are as follows. First, the $p=0$ calculations are equivalent to the ones of the standard MFT. In other words, our MOT may be understood as a general extension of MFT, and its systematic improvement is achieved by introducing the symmetric operator.
Second, the $p=0$ calculation does not capture any quantum entanglement because both the mean-operator and the trivial state are of the product forms. 
We emphasize that the mechanism of the improvement in MOT is  the entangled nature of the exchange interaction term in $\hat{S}_1$.
Third, our discussions may be easily generalized to a system with a larger group structure where MFT is well developed.   
Knowledge of the MFT may be used to choose a mean-operator.

{\it Case 2) Symmetry-protected-topological state : } 
Our MOT is applicable to topological states. One prime example is a symmetry-protected-topological (SPT) ground state of the cluster model under a transverse magnetic field in one spatial dimension\cite{Wen_QI}, 
\begin{eqnarray}
H_{2}(g) = - (1-g)\sum_{j=1}^{L-2} Z_j X_{j+1} Z_{j+2} - g \sum_{j=1}^L X_j. 
\label{eq:H2}
\end{eqnarray}
The $\mathbb{Z}_2 \times \mathbb{Z}_2$ symmetry with $\{ I, \prod_{e} X_j, \prod_{o} X_j, \prod_{e,o} X_j\}$ exists, where the subscripts $e$ and $o$ for even and odd sites. A trivial product state is $ |0\rangle =\prod_j |+\rangle_j$ with $X_j |+\rangle_j = |+\rangle_j$, which is the ground state of $H_2(g=1)$. 

We choose the operators, 
\begin{eqnarray}
&& \hat{M}_2(\phi_1,\phi_2 )=e^{-i  \phi_2 \sum_{j} Z_{j} Z_{j+1} }  e^{-i \phi_1 \sum_j Z_j} \nonumber \\
&&\hat{S}_2(\{\alpha, \beta\}_p) = \prod_{a=1}^p e^{-i \beta_a \sum_j  Z_j {X}_{j+1} Z_{j+1}} e^{-i \alpha_a \sum_j  {X}_j}, \nonumber
\end{eqnarray}
and the zero depth level ($p=0$) calculation gives the ground state energy,
\begin{eqnarray}
\epsilon^{(0)}(\phi_1,\phi_2)  = (1-g)  C_{2\phi_1} S_{2\phi_2}^2 -g C_{2\phi_1}  C_{2\phi_2}^2.
\end{eqnarray}
Minimizing the energy density, a quantum phase transition  is obtained at $g_c^{(0)}=1/2$, which is the same as the exact one \cite{Doherty09}. The trivial product state with $(\phi_1 = \phi_2=0)$ appears as the ground states in $1/2 <g<1 $, while an entangled state with $(\phi_{1}=\pi/2, \phi_{2}=\pm \pi/4)$ optimizes the model below $g<1/2$.

\begin{figure}[t]
\centering
\includegraphics[width=0.45\textwidth]{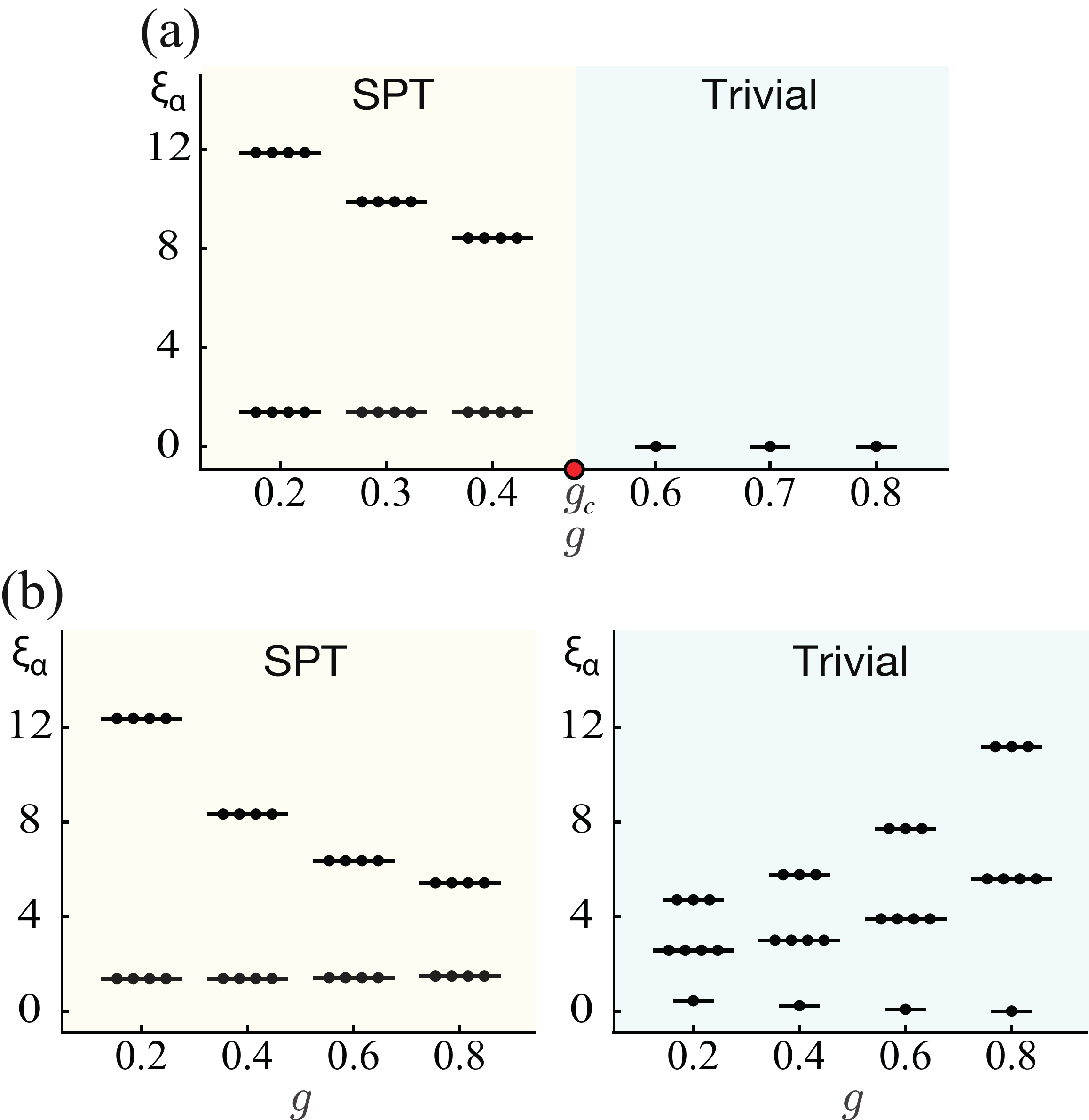}
 \caption{ (a) Entanglement spectrum, $\varepsilon_{\alpha} = e^{-\xi_\alpha}$, of $H_2(g)$ obtained by the depth-one MOT with four variational parameters. 
 The topological phase transition at $g_c$ is well captured by the number of degeneracy of the spectrum, where the four-fold degeneracy for $ g < g_c$ is the characteristic of the SPT phase.  
 (b) Entanglement spectrums of the depth-two QAOA states with four variational parameters. The initial states are chosen to be the ground states of $H_2(g = 0)$ (left) and $H_2(g = 1)$ (right), respectively. 
Numerical calculations are performed with 20 qubits, and the  periodic boundary condition is imposed.} 
  \label{fig:1}
\end{figure}

We stress that the mean-operator, $\hat{M}_2(\phi_1, \phi_2) $, is {\it entangled} because it cannot be written as a product of local operators unless $\phi_2=0$. 
The higher-order term $Z_j Z_{j+1}$ is essential and becomes the source of anomalous symmetry action at boundaries.   We present entanglement spectrums of our variational ground states with $p=1$ in Fig.\,\ref{fig:1}\,(a), where all levels exhibit the four-fold degeneracy below the critical value $g_c\simeq 0.56$ while the four-fold degeneracy disappears above $g_c$ \cite{Pollmann12}.  Note that the number of variational parameters are four. 
To demonstrate advantages of our MOT, we perform conventional QAOA calculations and illustrate their entanglement spectrums in Fig.\,\ref{fig:1}\,(b).
To use four variational parameters, the conventional depth-two QAOA are performed with different initial states.  
In the left (right) figure, the ground state of $H_2(g)$ with $g=0$ ($g=1$) is chosen as the initial state, respectively. 
The absence of quantum phase transitions is manifest, indicating that the MOT outperforms the conventional QAOA. 

Note that  the mean operator breaks $\mathbb{G}$ for generic values of $(\phi_1, \phi_2)$, but the one with $(\pi/2, \pi/4)$ produces a symmetric state. 
Therefore, the ground state with the mean-operator becomes not only entangled but also symmetric under $\mathbb{G}$. 
This is equivalent to the well-known fact that a SPT state cannot be obtained by acting local symmetric transformations on a trivial state.

 {\it  Case 3) Competition physics: } 
 One significant advantage of our MOT is the expressive power to describe both symmetry-broken and topological states, which allows us to access intriguing phenomena such as competition physics between spontaneously-symmetry-broken and topological states. 

As a proof-of-principle, we consider the Levin-Gu model \cite{Levin12} with a ferromagnetic exchange interaction term and a transverse-field term with two parameters ($g, \lambda$) on a triangular lattice, 
\begin{eqnarray}
H_3(\lambda, g)&=&(1-\lambda)H_{LG}+gH_{ex}+\lambda H_{PM}. 
\end{eqnarray}
The Levin-Gu Hamiltonian, trivial paramagnetic Hamiltonian, and the ferromagnetic exchange Hamiltonians are
\begin{eqnarray}
H_{LG} &=& -\sum_{j} B_j, \, \, H_{ex}=- \sum_{\langle jk \rangle}Z_jZ_k, \, \, H_{PM}=- \sum_{j} X_j, \nonumber 
\end{eqnarray}
with the dressed qubit term, $B_j=-X_j \prod_{\langle jkl \rangle}i^{(\frac{1-Z_k Z_l}{2})}$, where $\langle jkl \rangle$ is the nearest neighbor three sites around the site $j$. 
The Hamiltonian enjoys the $\mathbb{Z}_2$ symmetry, $\{ I, \prod_j X_j \}$, and a trivial product state is $ |0\rangle =\prod_j |+\rangle_j$ with $X_j |+\rangle_j = |+\rangle_j$, which is the ground state of $H(\infty, g)$ for a finite $g$.  Note that each term of the three Hamiltonians captures physics of symmetry-protected-topological, spontaneously-symmetry-broken, and trivial states, respectively. 

We choose the mean-operator, 
\begin{eqnarray}
\hat{M}_3(\theta,\phi, \psi) = \prod_{\langle jkl \rangle} e^{-i\theta Z_jZ_kZ_l-i\phi(Z_j+Z_k+Z_l)}  e^{-i \psi \sum_j Y_j}. \nonumber
\end{eqnarray}
Each operator associated with the three parameters ($\theta, \phi, \psi$) breaks the $\mathbb{Z}_2$ symmetry generically. The term with  $\theta$ is made of three adjacent qubits, describing non-trivial entanglement between adjacent qubits. For simplicity, a symmetric operator is set to be the identity in this work. It is straightforward to increase the depth, which may be used to properties of a refined many-body wavefunction. For example, one can show that the behaviors of the strange correlator \cite{Xu2014} persist around the depth-zero calculations perturbatively. 

\begin{figure}[tb]
\centering
\includegraphics[width=0.48\textwidth]{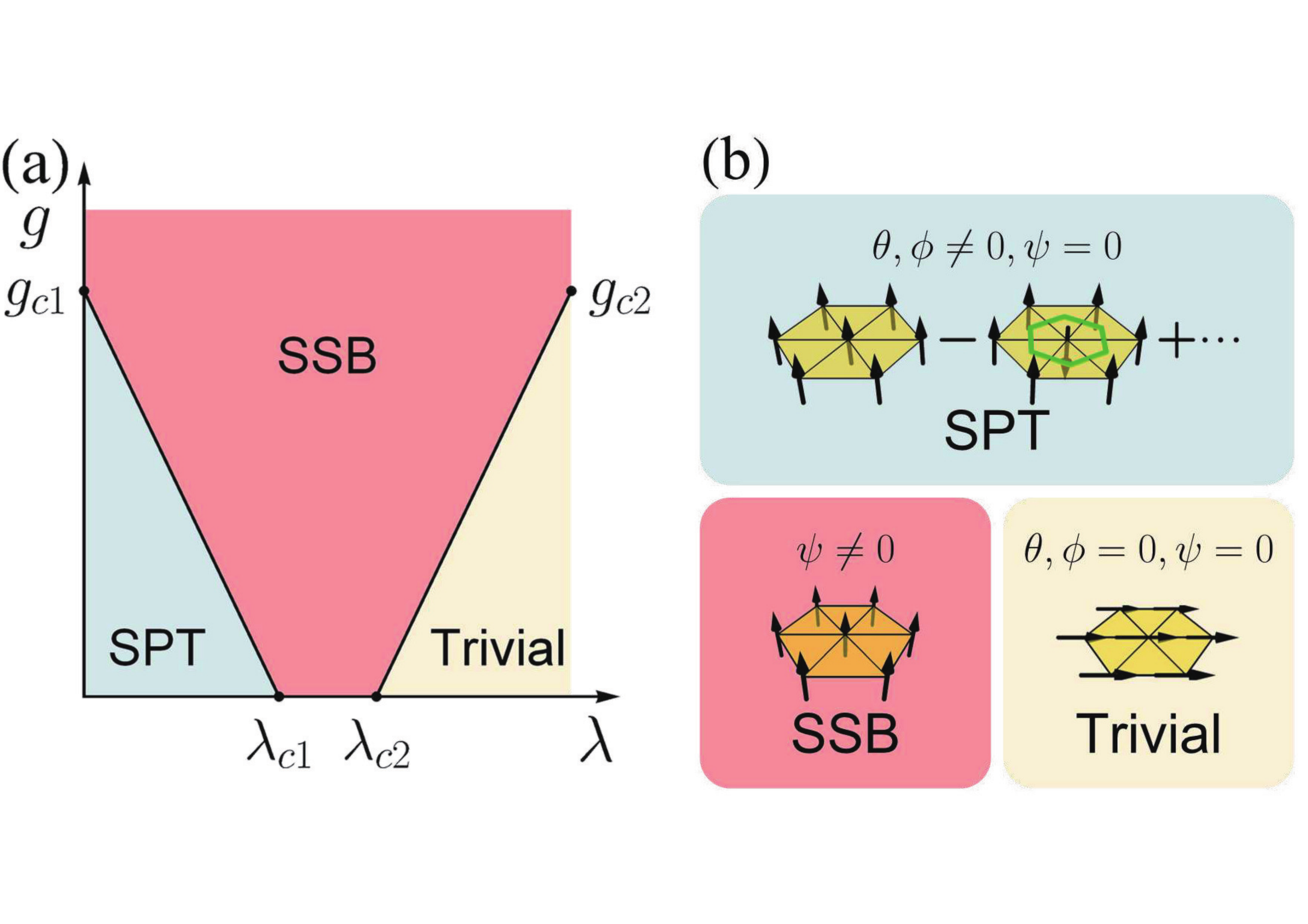}
  \caption{Competition between symmetry-protected-topological  and spontaneously-symmetry-broken (SSB) states. (a) The depth-zero MOT phase diagram of $H_3(\lambda, g)$. The critical values are $\lambda_{c1} =0.4$, $\lambda_{c2} =0.6$, $g_{c1} =g_{c2}=1/6$. The trivial and SPT states are characterized by $(\theta, \phi, \psi )=(0,0,0), (-\pi/8, \pi/24,0)$, respectively, and the spontaneously-symmetry-broken state is characterized by $\psi \neq 0$. 
  (b) Illustration of the wave-functions of the three states.}
 \label{fig:4}
\end{figure}

Varying with ($\theta, \phi, \psi$), we find three different phases in the $\lambda$-$g$ plane, as shown in Fig.\,\ref{fig:4}. We stress that the spontaneously-symmetry-broken state exists even without the exchange interaction ($g=0$), and its region becomes larger with a bigger $g$.
This clearly shows competition physics between symmetry-protected-topological and spontaneously-symmetry-broken states. 
It is an intriguing open question whether the spontaneously-symmetry-broken state survives for calculations with a non-zero depth, which we leave for future works. 
Also, another scenario associated with a stripe-order is recently suggested \cite{Scaffidi2020}, which may also be interesting to test in the future.    

Our model has the non-trivial duality ($\lambda \leftrightarrow 1-\lambda$), which makes the phase boundary symmetric in  the $\lambda$-$g$ plane (see SI).  
The quantum phase transitions between spontaneously-symmetry-broken and symmetry-protected-topological states and the ones between spontaneously-symmetry-broken and trivial states are continuous, and we argue that both of them are in the Ising universality class based on the duality. 
Though they are in the same universality class in the thermodynamic limit, it is interesting how their edge modes behave differently at the transitions.

{\it Discussion : } 
Our MOT overcomes the fundamental limitations of QAOA and enables to prepare a target state with a shallow circuit. A mean-operator  captures essential characteristics of the target state, and thus finding a mean-operator is at the heart of MOT. 
Intuitions from condensed matter theory such as symmetry and entanglement properties can be particularly useful, as shown the above cases where the non-trivial property under a given symmetry and the degree of entanglement of an operator are two important factors.  
Investigation of a class of mean-operators may provide an alternative way to classify quantum-many-body states.

Our MOT is applicable to near-term quantum simulations. The hybrid algorithms have already been demonstrated in the one-dimensional trapped ion problems \cite{Zoller19}, and its possibilities in a few qubits have been proposed \cite{Hsieh2019, Lukin2020}. Since our MOT is a natural extension of the hybrid algorithms, we believe that applications of MOT are plausible and powerful, especially in 2D and 3D. 

We propose that preparations of SPT states in quantum simulators of $^{87}$Rb neutral atoms or trapped $^{40}$Ca$^+$ ions may be particularly interesting. A realization of the mean-operators for each SPT state is a key step, and our results, for example $\hat{M}_2(\phi_1,\phi_2 )$, may be a useful guideline because an exchange interaction between adjacent qubits is a dominant Hamiltonian term in both $^{87}$Rb neutral atoms or trapped $^{40}$Ca$^+$ ions. Since mean-operators may depend on microscopic properties, we leave specific forms of mean-operators with propoer initial states for future works.

We anticipate that our MOT can be extended and applied in a variety of ways. For instance, one can introduce the spatial fluctuations of the mean-operator that may provide new insight into quantum field theories. Furthermore, a mean-operator may be extended to be non-unitary, and then the MOT can access topologically ordered states, similar to the recent  realization of toric code state \cite{Google2021} (see SI also).
Our MOT may be implemented for studying exotic interacting systems, including fractons \cite{fractons} and non-Fermi liquid states, which we leave for future works.

{\it Acknowledgements : } 
This work was supported by National Research Foundation of Korea under the grant numbers NRF-2019M3E4A1080411, NRF-2020R1A4A3079707, NRF-2021R1A2C4001847 (DK, PN, EGM) and NRF-2020R1I1A3074769 (HYL).

\bibliographystyle{apsrev}
\bibliography{references.bib}

\clearpage
\onecolumngrid
\begin{center}
\textbf{\large Supplementary Material for \\``Advancing Hybrid Quantum-Classical Algorithms via Mean-Operators''} 
\end{center}

\setcounter{equation}{0}
\setcounter{figure}{0}
\setcounter{table}{0}
\setcounter{page}{1}

\section{1. Comparison between Mean-Field Theory and Mean-Operator Theory}
Let us recall the standard approach of MFT in  the transverse field Ising model in $d$ spatial dimensions, 
\begin{eqnarray}
H_1(g) = - \sum_{\langle i,j \rangle} Z_i Z_j - g \sum_j X_j, 
\end{eqnarray}
where Pauli matrices $X_j,Y_j,Z_j$ at site $j$ are used with a dimensionless coupling constant, $g$.

The conventional MFT starts with introducing a mean-field Hamiltonian, 
\begin{eqnarray}
H_{1,MF}(g) &=&  \sum_j h_j, \quad h_j = -2d m  Z_j -g X_j+d m^2, \nonumber \\
 &=&  \sum_j \big( -\sqrt{g^2 + 4d^2 m^2} e^{-i \phi Y_j} X_j e^{+i \phi Y_j}+d m^2 \big), \nonumber
\end{eqnarray} 
where the order parameter and energy density are
\begin{eqnarray}
m= \langle G_{MF} | Z_j | G_{MF} \rangle, \quad \epsilon_G^0= \frac{E_G^0}{N} = - \sqrt{g^2 + 4d^2m^2} +d m^2. \nonumber 
\end{eqnarray}
Solving the MF Hamiltonian, its ground state is
\begin{eqnarray}
| G_{MF} \rangle = e^{-i \phi_0 \sum_j Y_j} |0 \rangle, \quad \tan(2\phi_0) = \frac{2d m}{g}. 
\end{eqnarray}
The stationary condition 
gives
\begin{eqnarray}
m_0 =\pm \sqrt{1-\frac{g^2}{4 d^2}}, \nonumber 
\end{eqnarray}
and the mean-field critical point is $g_{c}^{(0)} = 2 d$. 

The MFT results are connected with MOT by choosing the operators, 
\begin{eqnarray}
\hat{M}(\phi) = e^{-i \phi \sum_j Y_j}, \quad \hat{S}=I,
\end{eqnarray}
which manifest the equivalence between MFT and the lowest order calculation of MOT.

\section{ 2. Finite-depth MOT }
\begin{figure}[!b]	\includegraphics[width=0.5\textwidth]{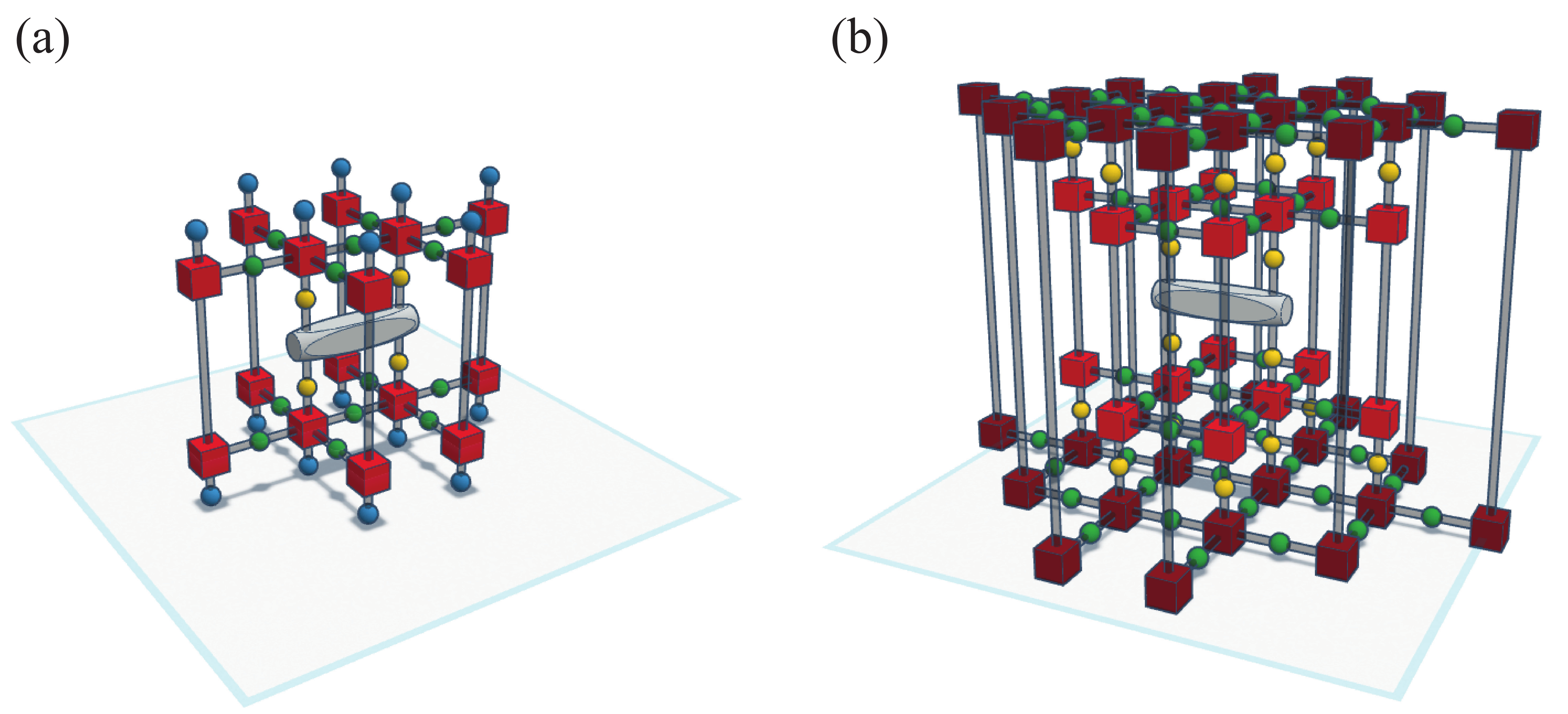}
	\caption{ Tensor network representations for the expectation value of a two-site operator  with respect to (a) the depth-one and (b) depth-two MTO. }
	\label{fig:2d_depth_tn}
\end{figure}
We employ the tensor network\,(TN) representations of the wavefunctions and various operators for a finite-depth MOT. 
For example, the trivial product state $|0\rangle$ can be recast as a bond dimension $D=1$ TN state regardless of the spatial dimension. Similarly, the tensor product of the local unitary gates, e.g., the mean-operator $\hat{M}_1(\phi) = \prod_i e^{i\phi Y_i}$,  is also represented as a $D=1$ TN operator. On the other hand, the product of the two-site or multi-site unitary gates requires a bond dimension larger than $1$, which indicates the generation of quantum entanglement. To be more concrete, the symmetric operator for SSB, $\hat{S}_1$, includes $\prod_{\langle ij \rangle} e^{i \beta Z_i Z_j }$ that can be recast as 
\begin{align}
	\prod_{\langle ij \rangle} e^{i \beta Z_i Z_j } = {\rm tTr} \left( \prod_i T \prod_{\langle ij\rangle} B \right),
\end{align}
where ${\rm tTr(\cdots)}$ stands for the tensor trace of the TN and $T$\,($B$) denotes a site\,(bond) tensor sitting on each vertex\,(link) in the lattice:
\begin{align}
	& [T]_{\alpha_1 \alpha_2 \cdots \alpha_z }^{ss'} 
	= \begin{cases}
		I_{ss'}\quad {\rm if}\quad \alpha_1 + \alpha_2 + \cdots + \alpha_z = {\rm even} \\
		Z_{ss'}\quad {\rm if}\quad \alpha_1 + \alpha_2 + \cdots + \alpha_z = {\rm odd}
	\end{cases},\nonumber\\
	& [B]_{\alpha_i \alpha_j }
	= \begin{pmatrix}
		\cos\beta & 0 \\
		0 & -i\sin\beta
	\end{pmatrix}_{\alpha_i \alpha_j },
\end{align}
where $z$ is the coordination number. 

We stress that MOT is particularly powerful if mean-operators and symmetric operators are unitary. 
It is because only local degrees of freedom are necessary to evaluate the variational energy and local observables whose size depends on the depth of employed operators. For example, the expectation value $\langle Z_i Z_j \rangle$ f or the $p=1$ MOT for the two-dimensional transverse field Ising model on the square lattice is illustrated in Fig.\,\ref{fig:2d_depth_tn}\,(a). Here, the blue spheres on top and bottom stand for $ e^{i\phi Y} |+\rangle$ and its complex conjugate, and the red cube and green sphere for the site and bond tensors defined above. The yellow sphere denotes $e^{-i\alpha Z}$, and the two-site operator is $Z_i Z_j$. Further, the size of the cluster increases with the depth, and thus the TN becomes larger as presented in Fig.\,\ref{fig:2d_depth_tn}\,(b) that represents the $p=2$ expression for $\langle Z_i Z_j \rangle$. Its higher-dimensional generalization is straightforward.

\begin{figure}[!t]	\includegraphics[width=0.5\textwidth]{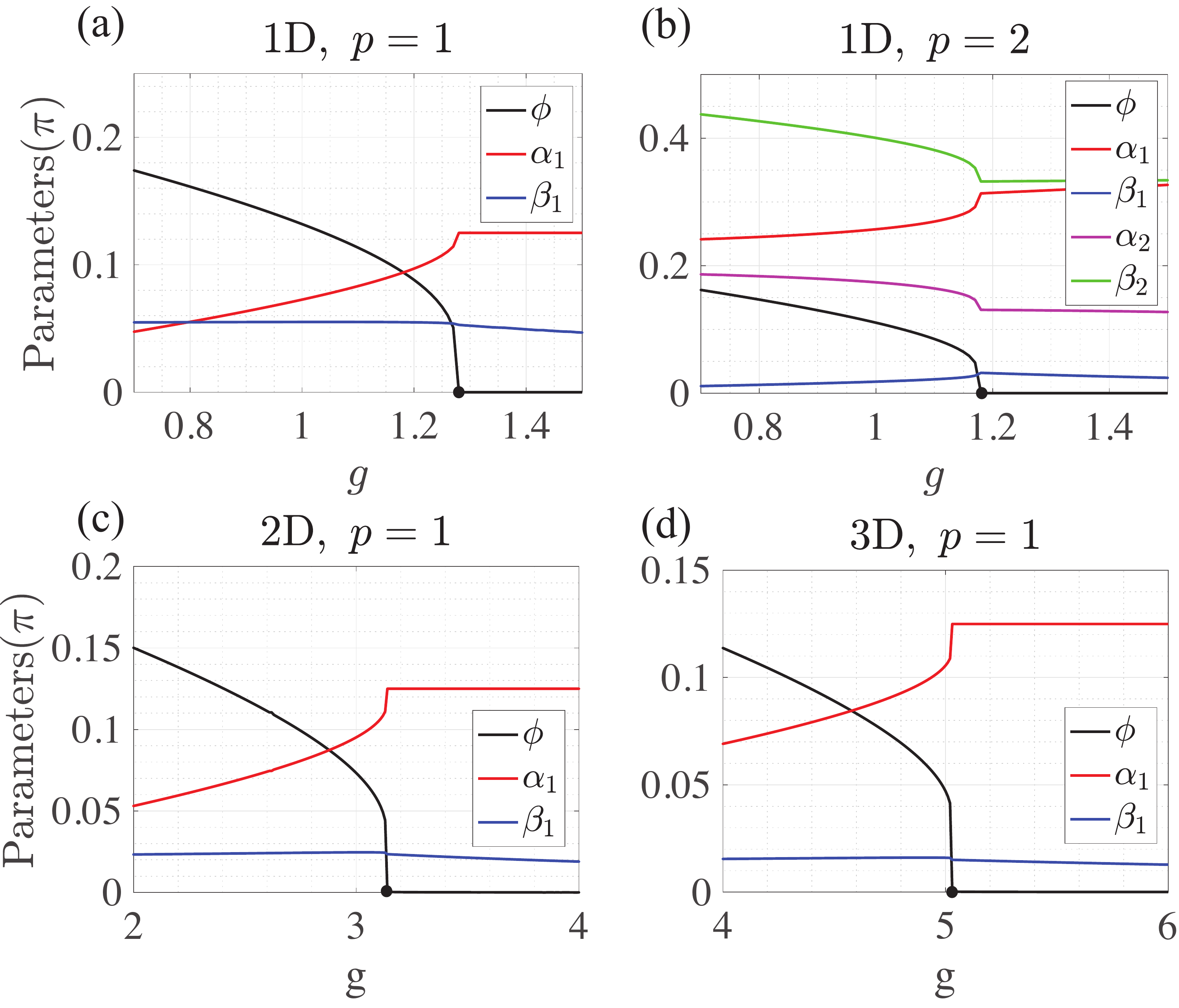}
	\caption{ Optimized variational parameters in the MOT for the (a)-(b) one-dimensional, (c) two-dimensional, and (d) three-dimensional transverse field Ising model defined in Eq.\,(2) in the main text. }
	\label{fig:ssb_var_param}
\end{figure}

The TN representations allow obtaining the minimum of the variational energy conveniently, varying with the parameters $\phi$ and $\{\alpha_i, \beta_i\}$. Fig.\,\ref{fig:ssb_var_param} presents the optimized parameters for the transverse field Ising model in one, two, three spatial dimensions. Notice that the variational parameter ($\phi$) of the mean operator $\hat{M}$ plays the role of a mean-field order parameter as it should be.

\section{ 3. Symmetry Protected Topological Phase}
\begin{figure}[t]
\includegraphics[width=0.32\textwidth]{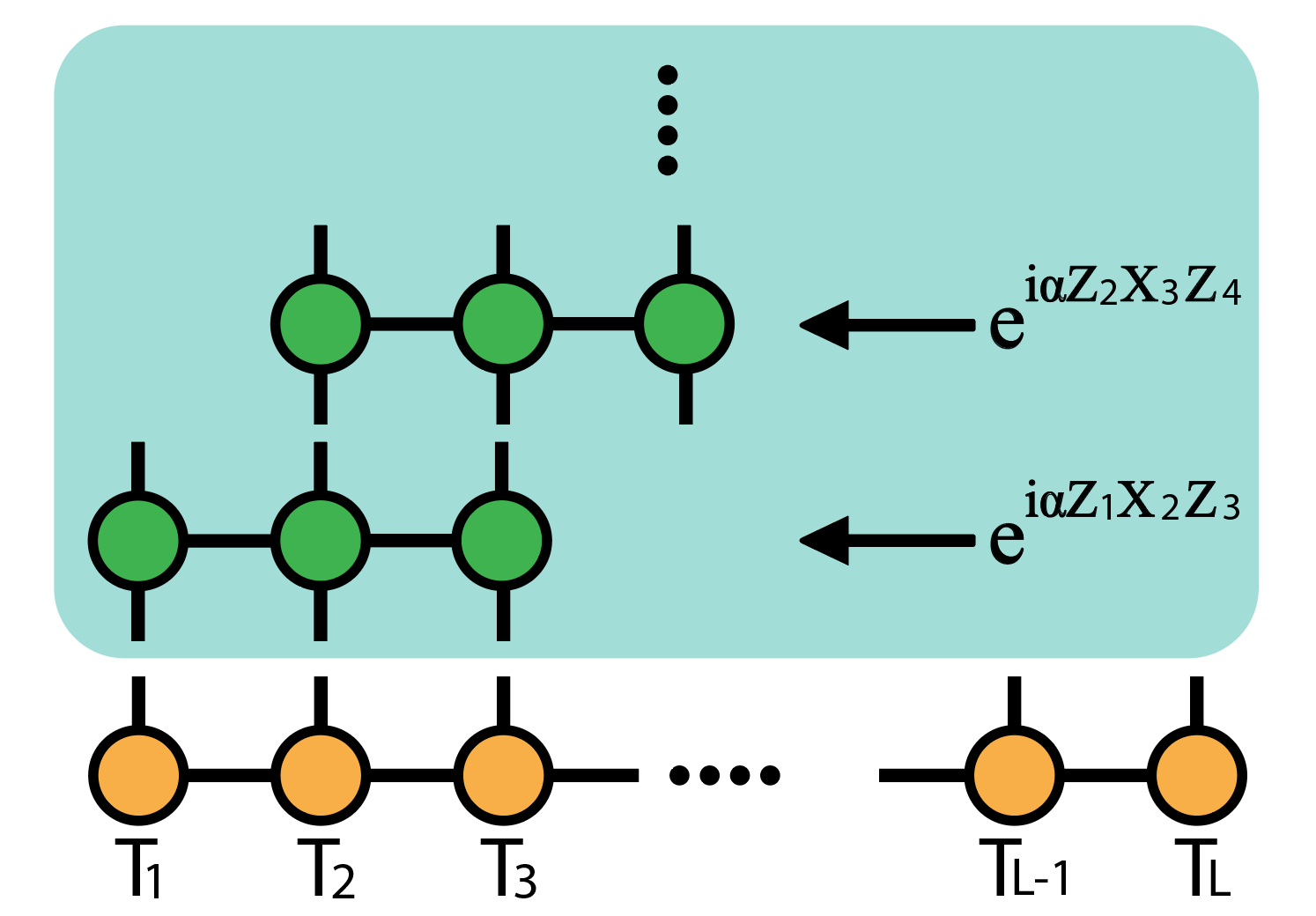} 
  \caption{Applying $e^{i\alpha Z_{i}X_{i+1}Z_{i+2}}$ to the MPS by using each $e^{i\alpha Z_{i}X_{i+1}Z_{i+2}}$. $T_i$ is each site tensor of the MPS.}
 \label{fig:10}
\end{figure}

We employ the MPS and MPO representations for the wavefunction ansatz and the symmetric operators\,(or the time-evolution unitary operators) to study the Hamiltonian $H_2$ exhibiting the transition between trivial and SPT phases. As highlighted previously, the tensor network representation lowers the computational costs significantly for evaluating the variational energy and thus enables us to treat deeper MOTs. In particular, in this section, we provide detailed information on the MPS and MPO. The initial state $|\{+\} \rangle$ can be expressed by the following MPS representation:

\begin{align}
	|\{+\}\rangle = \sum_{\{\sigma_i\}} {\rm Tr}  \left( A^{\sigma_1} \cdots A^{\sigma_L} \right) |\sigma_1 \cdots \sigma_L\rangle ,
\end{align}
with $A^{\sigma} = \frac{1}{\sqrt{2}}I$. Here, $I$ stands for the $(2\times2)$-identity matrix. The unitary operators can be recast as

\begin{align}
	U = \sum_{\{\sigma_i\}} {\rm Tr}  \left( T^{\sigma_1}_{\sigma_1'} \cdots T^{\sigma_L}_{\sigma_L'} \right) |\sigma_1 \cdots \sigma_L\rangle \langle\sigma_1' \cdots \sigma_L' |,
\end{align}
where 
\begin{align}
	T^{\sigma}_{\sigma'} = 
	\begin{cases}
		\cos\alpha\, [I]^{\sigma}_{\sigma'} + i\sin\alpha\, [Z]^{\sigma}_{\sigma'} \quad {\rm for}\quad U = e^{i\alpha\sum_i Z_i}\\		
		\cos\alpha\, [I]^{\sigma}_{\sigma'} + i\sin\alpha\, [X]^{\sigma}_{\sigma'} \quad {\rm for}\quad U = e^{i\alpha\sum_i X_i}
	\end{cases},
\end{align}
and
\begin{align}
	&T^{\sigma}_{\sigma'} = \nonumber\\
	&\left(\begin{array}{cc} \cosh(\alpha)[I]^{\sigma}_{\sigma'} & \sqrt{\cosh(i\alpha)}\sqrt{\sinh(i\alpha)}[Z]^{\sigma}_{\sigma'}\\ \sqrt{\cosh(i\alpha)}\sqrt{\sinh(i\alpha)}[Z]^{\sigma}_{\sigma'} & \cosh(i\alpha)[I]^{\sigma}_{\sigma'} \end{array}\right)
\end{align}
for $U = e^{i\alpha\sum_i Z_i Z_{i+1}}$.
The time-evolution operator $e^{i\alpha\sum_{i}Z_{i}X_{i+1}Z_{i+2}}$ is represented as a product of local MPOs as illustrated Fig.\,\ref{fig:10}. The three-site MPO of $e^{i\alpha Z_{i}X_{i+1}Z_{i+2}}$ is written in the following way:
\begin{align}
	\left({\begin{array}{cc} \cos(\alpha)I_{i} & i \sin(\alpha)Z_{i} \end{array}}\right) \left(\begin{array}{cc} I_{i+1} & 0\\ 0 & X_{i+1} \end{array}\right)\left(\begin{array}{c} I_{i+2} \\ Z_{i+2} \end{array}\right).
\end{align}

In order to obtain the entanglement entropy, we exploit the canonical form of the ansatz by following the standard procedure.

\section{4. Depth-zero calculations of $H(g, \lambda)$}

We provide detailed information on the zero-depth calculations of $H(g,\lambda)$. The Hamiltonian enjoys the duality,
\begin{eqnarray}
H(1-\lambda, g) = \Big( \hat{M}(-\frac{\pi}{8}, \frac{\pi}{24},0)^{\dagger} \Big)  H(\lambda, g)  \Big(\hat{M}(-\frac{\pi}{8}, \frac{\pi}{24},0) \Big), \nonumber
\end{eqnarray}
generalized from the original Levin-Gu model. 
The variational state with  the mean operator, $\hat{M}(\theta,\phi,\psi)$, gives the variational energy,
\begin{eqnarray}
 &&  E_G(\theta, \phi,\psi;\lambda,g) \equiv\left \langle \theta, \phi,\psi\left|H(\lambda,g)\right|\theta, \phi, \psi\right \rangle \nonumber \\
    &&=(1-\lambda)\left[S_{12\phi}F_1\left(\frac{\pi}{4}-2\theta,2\psi \right)-C_{12\phi}F_2\left(\frac{\pi}{4}-2\theta,2\psi \right) \right] \nonumber \\
   &&-\lambda \left[ C_{12\phi} F_1\left(-2\theta,2\psi \right)+S_{12\phi}F_2\left(-2\theta,2\psi \right) \right]-3gS^2_{2\psi}, \nonumber \label{result}
\end{eqnarray}
 with
\begin{eqnarray}
F_1(\theta,\psi)&\equiv& \frac{1}{4} C_{\psi}\Big(C_{2\theta}(3+C_{2\theta}^2)-C_{2\theta}S_{2\theta}^2(6S_{\psi}^2+9S_{\psi}^4) \Big),  \nonumber \\
F_2(\theta,\psi)&\equiv& \frac{1}{8}C_{\psi}\Big(6S_{2\theta}(1+3C_{2\theta}^2)S_{\psi}^2-S_{2\theta}^3(12S_{\psi}^4+2S_{\psi}^6) \Big). \nonumber 
\end{eqnarray}
The variational energy satisfies useful properties including
\begin{eqnarray}
E_G(\theta,\phi,\psi;\lambda,g)&=&E_G(\theta+\frac{\pi}{2},\phi,\psi;\lambda,g), \nonumber \\
E_G(\theta,\phi,\psi;\lambda,g)&=&E_G(\theta,\phi+\frac{\pi}{6},\psi;\lambda,g),\nonumber \\
E_G(\theta,\phi,\psi;\lambda,g)&=&E_G(\theta+\frac{\pi}{4},\phi+\frac{\pi}{12},\psi;\lambda,g), \nonumber \\
E_G(\theta,\phi,\psi;\lambda,g)&=&E_G(\frac{3\pi}{8}-\theta,\frac{\pi}{24}-\phi,\psi;1-\lambda,g), \nonumber 
\end{eqnarray}
which allow us to restrict the range of the parameters. 
The phase boundaries may be determined by using the properties. For example, the boundary between the trivial and SSB phases is determined by the condition, $E_G(0,0,0;\lambda,g) = E_G(0,0,\psi;\lambda,g)$, near the onset of SSB. With a small parameter $\psi$, the equation becomes
\begin{eqnarray}
-\lambda=(1-\lambda)(-3\psi^2)-\lambda (1-2\psi^2)-12g\psi^2 +O(\psi^3), \nonumber
\end{eqnarray}
for $\lambda \in [0.5,1]$. Thus, its boundary is 
\begin{eqnarray}
g=\frac{5}{12}\lambda-\frac{3}{12} \label{boundary1},
\end{eqnarray}
and the duality enforces another phase boundary between SPT and SSB,  
\begin{align}
g=-\frac{5}{12}\lambda+\frac{2}{12} \label{boundary2},
\end{align}
for $\lambda \in [0,0.5]$.

\section{5. mean-operator theory of $Z_2$ gauge theory}

In this section, we show that our MOT can be employed to construct a topologically ordered state, if one allows a non-unitary mean-operator. As an illustrative exmaple, we consider the $\mathbb{Z}_2$ gauge theory in a $d$-dimensional hypercubic lattice\,($d \ge 2$) \cite{Sachdev_topo}, 
\begin{align}
	& H_{3} (g) = - \sum_{a^*}  F_{a^*} - g \sum_{(ij)} X_{(ij)}, \nonumber\\
	& F_{a^*} \equiv \prod_{(ij) \in a^*} Z_{(ij)}.  	
\end{align}
Qubits with Pauli matrices on  each link $(ij)$ connecting sites $i$ and $j$ are considered, and the flux operator\,($F_{a^*}$) is defined on a dual lattice labeled as $a^*$. 
The Hamiltonian enjoys the local $\mathbb{Z}_2$ symmetry, whose element is a product of $X$'s on the links sharing the site $i$, $G_{i} =\prod_{ (kl) \in i} X_{(kl)} $. A symmetric product state is $ |0\rangle =\prod_{(ij)} |+\rangle_{(ij)}$ with $X_{(ij)} |+\rangle_{(ij)} = |+\rangle_{(ij)}$, which is the ground state of $H_3(\infty)$.  

\begin{figure}[!t]
\centering
\includegraphics[width=0.45\textwidth]{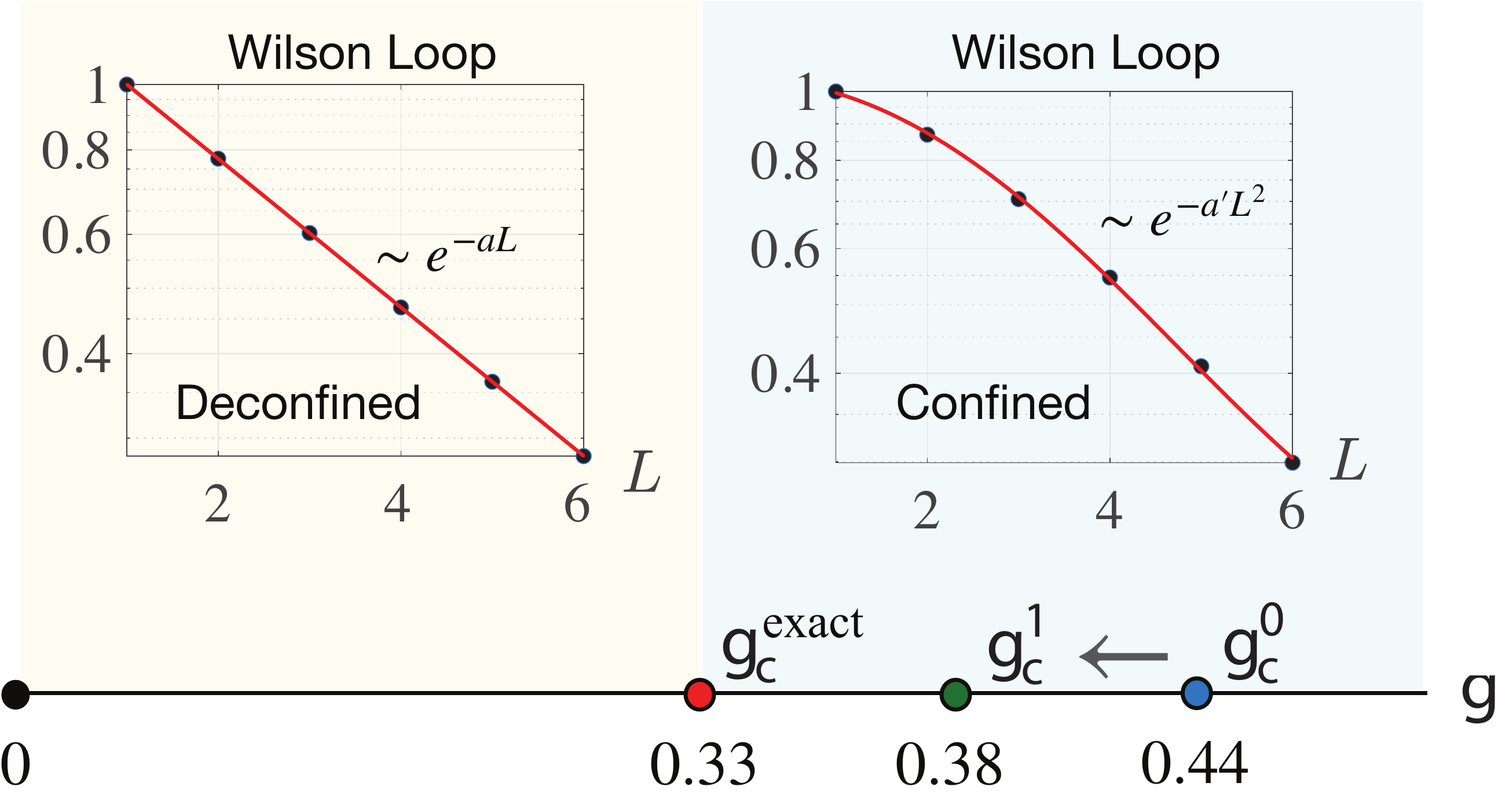}
 \caption{ MOT Phase diagram of the $\mathbb{Z}_2$ gauge theory, $H_3(g)$, in two spatial dimensions. Here, $g_c^{\rm exact}\simeq 0.33$ \cite{Blote02} is the true critical point, while $g_c^{(0)}$ and $g_c^{(1)}$ denote the transition points estimated by the depth-zero and depth-one MOT, respectively. The expectation values of the Wilson loop operator are illustrated to distinguish the two phases with the perimeter and area laws. }
 \label{fig:z2_theory}
\end{figure}

We choose the operators, 
\begin{eqnarray}
&&\hat{M}_3(\phi)= {\rm P}_G  \cdot e^{-i \phi \sum_{(ij)} Y_{(ij)}} \nonumber \\
&&\hat{S}_3(\{\alpha, \beta\}_p) = \prod_{a=1}^p e^{-i \alpha_a \sum_{(ij)} X_{(ij)}} e^{-i \beta_a  \sum_{a^*} F_{a^*}}, 
\end{eqnarray}
with  the gauge-symmetrization operator, $ {\rm P}_G \equiv \prod_i (1 + G_i)/2 $.  
Even at the depth-zero level, numerical evaluations are necessary, and the corner transfer matrix renormalization group with the tensor network representation is employed for measuring the variational energy and the observables, including the Wilson loop operator, $W = \prod_{\square} Z_j$, on a square $\square$. 
 
\begin{figure}[!t]	
	\includegraphics[width=0.5\textwidth]{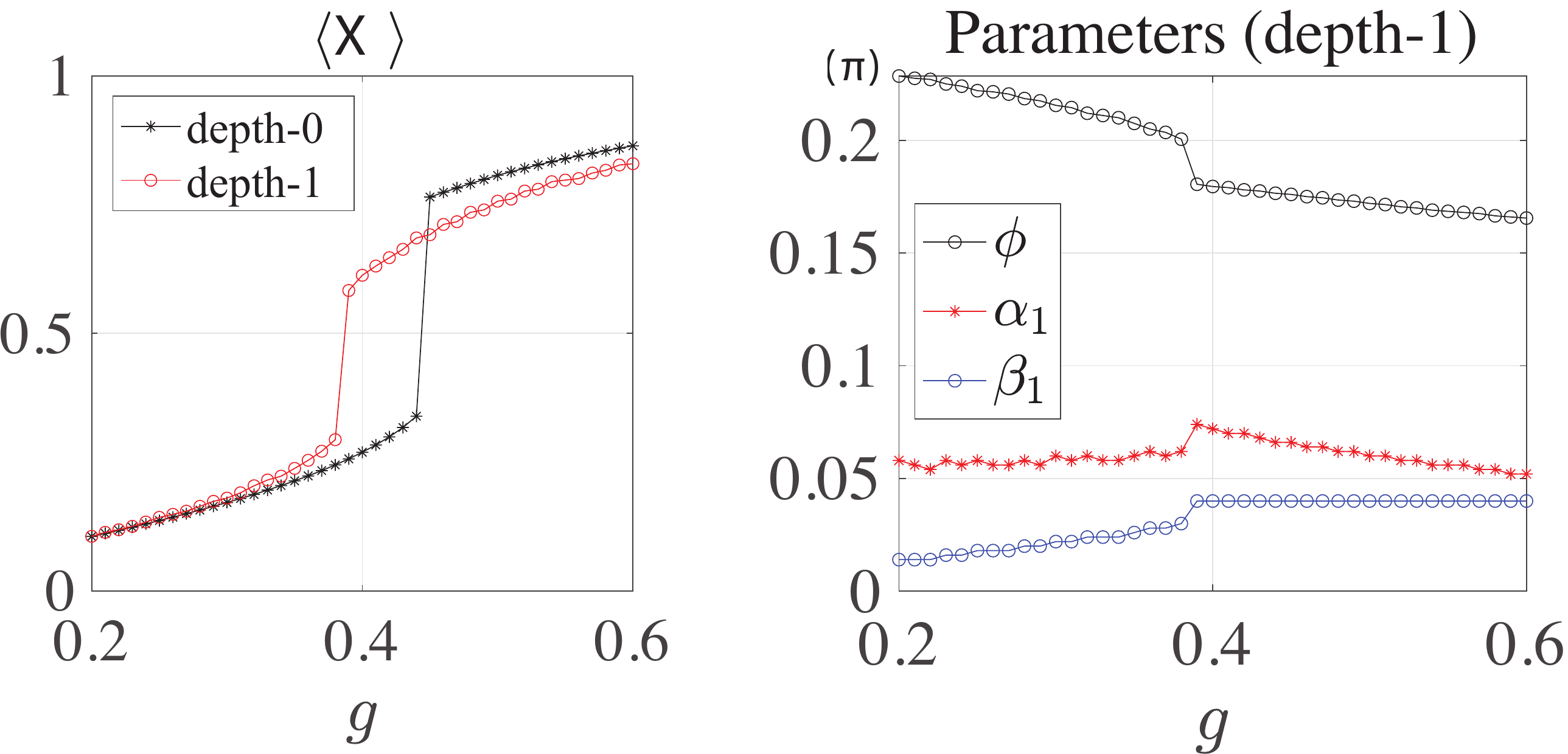}
	\caption{ Left: the expectation value of the Pauli operator $X$. Right: the optimized variational parameters of the depth-one MOT as a function of the coupling constant $g$ of the $Z_2$ gauge theory.}
	\label{fig:z2_params}
\end{figure}

We find two phases that exhibit the area and perimeter laws of the Wilson loop respectively as shown in Fig.\,\ref{fig:z2_theory}. The accuracy of the estimated critical point is improved significantly by increasing the depth from zero to one. We believe that the estimation becomes better and finally converges to the actual transition with a higher level depth as shown in Fig.\,\ref{fig:z2_params}.
The expectation value of the Pauli operator $X$ exhibits a discontinuity at the transition between the confined and deconfined phases as shown in the left panel of Fig.\,\ref{fig:z2_params}, meaning that the $p=0$ and $p=1$ MOT anticipate the first-order phase transition. However, the discontinuity becomes weaker while the transition point comes closer to the exact one $g_c \simeq 0.33$. We, therefore, believe that the MOT with sufficiently large depth may describe the continuous transition correctly. In order to optimize the variational parameters, we divide the parameter space\,($0\leq\phi,\alpha_1,\beta_1 \leq \pi/4$) equally and optimize to minimize the energy.  The optimized parameters for the $p=1$ MOT are presented in the right panel of Fig.\,\ref{fig:z2_params}. The uneven behavior of the optimized parameters is due to the finite distance between parameters.

We stress that  the mean-operator $\hat{M}_3(\phi)$ is not symmetric under the local $\mathbb{Z}_2$ transformation, but $\hat{M}_3(\phi) {\rm P}_{G}$ is symmetric for all $\phi$, indicating that the ansatz is guaranteed to be symmetric regardless of $\phi$.  
This property is crucial to explore the topologically ordered state because the action of the mean-operator opens a possibility to access a topologically distinct state while its symmetry remains intact. In a sense, it is similar to the case of spontaneously-symmetry-broken states where the action of a mean-operator accesses a symmetry-broken state.  
Strictly speaking, our choice of $\hat{M}_3(\phi)$ requires the number of quantum operations proportional to a system size similar to the recent realization of a topological order in a superconducting system with about twenty qubits \cite{Google2021}, which manifests the Lieb-Robinson bound with local unitary circuits \cite{Hsieh2019}. It is highly desirable to find a non-local mean-operator with a reduced number of quantum operators, and we leave this important matter for future works.

\end{document}